\documentclass[12pt]{JHEP3}

\pdfoutput=1
\usepackage{bm}
\usepackage{epsfig}
\usepackage{citesort}
\usepackage{graphicx}
\usepackage{multirow}

\def\be{\begin{equation}}
  \def\ee{\end{equation}}
\def\bea{\begin{eqnarray}}
  \def\eea{\end{eqnarray}}

\title{Searching for systematics in SNIa and galaxy cluster data using the cosmic duality relation }

\author{Arman Shafieloo\\
       Asia Pacific Center for Theoretical Physics, Pohang, Gyeongbuk 790-784, Korea\\
       Department of Physics, POSTECH, Pohang, Gyeongbuk 790-784, Korea \\
	Institute for the Early Universe, Ewha W. University, Seoul, 120-750, Korea\\ 
        E-mail: \email{arman@apctp.org}}

\author{Subhabrata Majumdar \\
Tata Institute for Foundamental Research, Mumbai, India\\
E-mail:\email{subha@tifr.res.in}}

\author{Varun Sahni\\
  Inter-University Centre for Astronomy and Astrophysics,
  Post Bag 4, Ganeshkhind, Pune 411~007, India \\
	E-mail: \email{varun@iucaa.ernet.in}}

\author{Alexei A. Starobinsky\\
Landau Institute for Theoretical Physics RAS, Moscow 119334, Russia\\
Research Center for the Early Universe (RESCEU), Graduate School of Science,
The University of Tokyo, Tokyo 113-0033, Japan \\
E-mail: \email{alstar@landau.ac.ru}}

     
\abstract{We compare two different probes of the expansion history of the universe, namely, 
luminosity distances from type Ia supernovae and angular diameter distances from galaxy clusters, using the Bayesian interpretation of Crossing statistic \cite{Crossing,Crossing_B} in conjunction with the assumption of cosmic duality relation. 
Our analysis is conducted independently of any a-priori assumptions
about the nature of dark energy. The model independent method which we invoke searches
for inconsistencies between SNIa and galaxy cluster data sets.
If detected such an inconsistency would imply the presence of systematics in either of
the two data sets.
Simulating observations based on expected WFIRST supernovae data and X-ray eROSITA + SZ Planck cluster data, we show that our method allows one to detect systematics with high precision and without advancing any hypothesis about the nature of dark energy.}

\begin{document}

\section{Introduction}                        
\label{sec:introduction}

A remarkable property of our universe is that is seems to be accelerating during the present epoch \cite{acceleration}. The precise cause of cosmic acceleration is presently unknown but might rest
in the presence of the cosmological constant or some other form of
`Dark Energy' (DE) capable of violating the strong energy condition.
Alternatively, cosmic acceleration might arise due to modifications to the
gravitational sector of the theory, such as $f(R)$ gravity, Braneworld models,
etc. However, as the simple example of a non-minimally coupled
scalar field shows, these two possibilities are only extreme particular cases of a more general notion of DE both including new non-gravitational physical fields and modifying gravity, too. Many of these possibilities have been extensively examined in recent years
\cite{DE_review} but firm conclusions as to the nature of dark energy
are still to be drawn.
One reason for this is that current observational data sets, despite steady
improvement, are still hampered by uncertainties both of a statistical
as well as systematic nature \cite{systematics}.

Indeed, in order for firm and robust conclusions to be drawn about the nature of
DE one will need to (a) minimize statistical uncertainties by increasing the
depth and quality of observational data sets,
(b) understand (and model better) the nature of systematics in the different kinds
of data sets used to explore cosmic acceleration.

The main evidence for cosmic acceleration currently comes from two types of
data sets: 

(i) Those probing the luminosity distance, $d_L$, by observing the flux, $\cal F$, of type Ia supernovae of given luminosity $L$ through
\be
{\cal F} = \frac{L}{4\pi d_L^2}~,
\ee
where
\be
\frac{H_0 d_L}{1+z} = \int\frac{dz}{h(z)}~, \hspace{10 mm} h(z)=\frac{H(z)}{H_0},
\label{eq:lum_dis}
\ee

in a spatially flat universe where $H(t)=\frac{\dot a}{a}$ and $a(t)$ is the FRW scale factor.

(ii) Those based on the angular diameter distance, $d_A$ to a source of spatial size $d$ via the relation
\be
\Delta \theta = \frac{d}{d_A}~.
\ee

Remarkably, for a wide range of cosmological models, the two distances are
related through the {\em cosmic duality relation} (CDR)
\be
d_L = (1+z)^2 d_A~.
\label{eq:duality}
\ee

In recent years numerous studies have devoted themselves to establishing whether or not the
CDR holds in practice \cite{duality}. The reason for this largely stems from the hope that
a violation of the equality in (\ref{eq:duality}) might signal the presence of new physics.
Such a violation may occur, for instance, through photon number non-conservation
either through axion-photon mixing \cite{csaki}, or because of photon absorption enroute to
the observer \cite{ChenKantowski09a}, or due to an incorrect modelling of the
 ultra-narrow beams from point sources such as type Ia supernovae \cite{Clarkson2012}.

However, since (\ref{eq:duality}) follows simply: (i) from the requirement that
sources and observers be connected via null geodesics in a Riemannian
space-time, (ii) the phase-space conservation of photons;
therefore the CDR remains valid for a very wide class of spatially homogeneous
(and even inhomogeneous) cosmologies \cite{eth33,bk04}.
Therefore it could well be that the cosmic duality relation is an {\em exact principle
in nature}. If this is indeed the case, then a violation of the CDR would no longer imply
new physics, but would signal instead to the presence of hitherto undetected systematics in
data sets relating either to $d_L$ or to $d_A$ (or to both).
Since the actual cosmological model of the present Universe is not known and there exist many models of 
dark energy which provide an alternative to an exact cosmological constant, it is interesting to investigate 
if one can compare different observational data and look for systematics in them without making any theoretical 
assumptions regarding the cosmological model.

The purpose of the present paper is to show how the presence of a systematic increase in the value of the cosmological distance modulus (as an observable)
\be
\mu \equiv m-M = 5\log_{10}{d_L} + 25
\label{eq:mu}
\ee

shows itself as an apparent violation of the cosmic duality relation.
 In this paper we use the idea of the Bayesian interpretation of Crossing Statistic \cite{Crossing,Crossing_B,Crossing_C} along with the smoothing method \cite{Shafieloo06,Shafieloo07,Shafieloo10} to compare cosmic distances from different data sets in a purely model independent manner.
We shall show that even a small change in $\mu$ can be detected
using the CDR and cosmological data sets which shall soon become available.
Using this approach we are able in fact to disclose differences between the two data sets without a need to make any assumptions about the nature of dark energy or to set any priors on cosmological parameters.


In the following, we first discuss the method that we use to compare different data sets. 
The smoothing method and the Bayesian interpretation of Crossing statistic will be discussed briefly and we
shall explain how we combine these methods to look for consistency between different data sets. This idea has been discussed briefly in \cite{Crossing_C} and in this work we explore the application of this method in greater detail
 and in a different context. In this approach we use the smoothing method together with the
 Crossing Statistic to reconstruct the cosmic distances from the two different datasets and set confidence limits on the reconstructed Crossing hyperparameters. In the absence of
 systematics in either of the data sets, the confidence limits of the Crossing hyperparameters should 
have a reasonable overlap. On the other hand, if the Crossing hyperparameters do not overlap nicely, 
then this would imply an inconsistency between data sets which could
 be interpreted as the existence of some sort of systematics. We shall
 apply our method on future simulated data where in one case we assume that there are no systematics in the data and in
the other we assume that there exists some form of systematics. Finally, we present our results
 and show how our method can discern the presence of systematics without any prior assumption on the 
cosmological model. 
We should note that the method we use in this paper can be generally used to compare different data sets in a model independent manner and this work is an application of the method to compare angular diameter distance data with luminosity distance data assuming the cosmic duality relation.

\section{Method and Analysis}                        
\label{meth}

\subsection{Smoothing Method}

The smoothing method is a completely model independent approach to derive the $d_L(z)$ relation directly from the data, without any assumptions other than the introduction of a smoothing scale. The only parameter used in the smoothing method is the smoothing width $\Delta$, which is constrained only by the quality and quantity of the data, and has nothing to do with any cosmological model. The smoothing method is an iterative procedure with each iteration typically giving a better fit to the data. It has been shown in \cite{Shafieloo06,Shafieloo07,Shafieloo10} that the final reconstructed results are independent of the assumed initial guess, $d_L(z_i)^g$. 

The modified smoothing method (error-sensitive) can be summarised by the following equation~\cite{Shafieloo10}:\\

\begin{eqnarray}
\label{eq:bg}
&&\ln d_L(z,\Delta)^{\rm s}=\ln
\ d_L(z)^g \nonumber\\
&& +N(z) \sum_i \frac{\left [ \ln d_L(z_i)- \ln
\ d_L(z_i)^g \right]}{\sigma^2_{d_L(z_i)}} 
\ {\rm exp} \left [- \frac{\ln^2 \left
( \frac{1+z_i}{1+z} \right ) }{2 \Delta^2} \right ],  \nonumber \\
&&N(z)^{-1}=\sum_i {\rm exp} \left
[- \frac{\ln^2 \left ( \frac{1+z_i}{1+z} \right ) }{2 \Delta^2} \right ] \frac{1}{\sigma^2_{d_L(z_i)}} ~,
\end{eqnarray}

\noindent where $d_L(z)$ is the data, $N(z)$ is the normalization factor, $d_L(z_i)^g$ is the initial guess model and $\Delta$ is the width of smoothing. 

The absolute brightness of type Ia supernovae is degenerate with $H_0$ since the observed quantity is the distance modulus $\mu(z)$. The outcome of the smoothing method is therefore $H_0d_L(z)/c\equiv d_L^{rec}(z)=(1+z)D(z)$. In this paper we set $\Delta=0.30$ which
 is similar to the value used in~\cite{Shafieloo10}. Complete explanation of the relations between $\Delta$, the number of data points, quality of the data and results of the reconstruction exercise can be found in~\cite{Shafieloo06,Shafieloo07}. It has earlier been shown that the smoothing method is a promising approach to reconstruct the expansion history of the universe. However, setting confidence limits has been an issue and in earlier work
the  bootstrap approach was used to set the confidence limits. In this paper, the reconstructed form of 
$d_L(z)$ will be used as a mean function in the full reconstruction process~\cite{Crossing_C}
which includes the idea of Bayesian interpretation of Crossing Statistic as explained in the next section.
 
\subsection{Reconstructing the Expansion History of the Universe using the Crossing Statistic }

The main idea behind the crossing statistic lies in the fact that the actual model of the universe and the 
reconstruction using smoothing would have one or two mild crossings: 
the distance modulus $\mu(z)_{\rm fiducial}$ of the fiducial cosmological model and the reconstructed 
$\mu(z)_{Smooth}$ 
would cross each other at one or two points in the redshift range defined by the data \cite{Crossing}. 

Furthermore, in a FRW universe the distance modulus monotonically increases with redshift. Consequently
any two cosmological models become virtually indistinguishable if the distance modulus of one of them 
is multiplied by a suitable function of the redshift. The coefficients of this function (multiplying $\mu$) constitute
the Crossing hyperparameters and the functions themselves will be called Crossing functions following 
\cite{Crossing_B}.
In our case the Crossing functions multiply the smoothed distance modulus reconstructed from the data
using the method described in the previous section. We shall refer to the smoothed functions as the
{\em mean} functions since they accurately describe the mean value of the function that is being smoothed,
which in this particular case is $\mu$. From the preceding argument, and that in \cite{Crossing_B}, it is clear that
the crossing functions, multiplied by the mean functions,
should virtually coincide with the actual model of the universe. \\

The reconstruction of the expansion history of the universe using Bayesian interpretation of Crossing statistic~\footnote{The Bayesian interpretation of Crossing Statistic is hidden in two prior assumptions that 1) Cosmic distances increases by redshift monotonically for all cosmological models hence there are no high frequency fluctuations in $\mu(z)$ and 2) since the distance modules of all cosmological models increases by redshift, $\mu(z)$ of any two cosmological models can become so close to each other at all redshifts up to an indistinguishable level if we multiply $\mu(z)$ of one of them to a suitable smooth mean function of degree $n$ with some particular values for the coefficients~\cite{Crossing_B}.} is therefore a combination of a non-parametric (smoothing) method with a parametric approach (using a Crossing function) to define and set the confidence limits on the cosmic expansion history. The crossing function is defined by 
Chebyshev polynomials~\cite{Crossing_B,Crossing_C}:

\begin{equation}
T_{II}(C_1,C_2,z)=1+C_1(\frac{z}{z_{max}})+C_2[2(\frac{z}{z_{max}})^2-1],
\label{eq:Cheb}
\end{equation}

and we fit 

\begin{equation}
\mu_{Smooth}^{T_{II}}(z)=\mu_{Smooth}(z) \times T_{II}(C_1,C_2,z)
\label{eq:main}
\end{equation}

\noindent to the data and find the best fit point $C^{best}_1,C^{best}_2$ in the hyperparameter space as well as the $C_1,C_2$ parameters related to the $1\sigma$, $2\sigma$ and $3\sigma$ confident limits. Each  
$T_{II}(C_1,C_2,z)$ (where $C_1,C_2$ pairs are within hyperparameter confidence contours) multiplied by
 $\mu_{Smooth}(z)$ represents a reconstruction of the expansion history of the universe consistent with the data.

Theoretically one  can use higher orders of Chebichev polynomial as well but this will result to more degrees of freedom and larger confidence limits which would eventually limit us from distinguishing between cases. It has been shown in~\cite{Crossing_B} that for the current and near future data, using up to second order of Chebishev polinomyals would be sufficient for the purpose of reconstruction. It may also look like that limiting the analysis up to using only second order of Chebichev polynomials might not make our method sensitive to the possible systematics of higher order but this is in fact true only if we use very smooth mean functions (usually suggeested by parameteric forms). In this work we use the well developed smoothing method to derive the mean function and smoothing method basically recover all detectable features of the data. The Crossing hyperparameters only generates smooth variations around this mean function. So if either of the datasets suffer from a kind of local systematic, lets say data is shifted up or down in a short range of redshift like a bump, this bump would be detected by the smoothing method and would be present in the mean function. In this paper we simulated the data only based on one type of systematics to show how the method works but for different kinds of systematics that affect the data in a similar way (changing the cosmic distances systematically) the method would be applicable. 

It has been shown that this method works very well in reconstructing the expansion history of the universe and 
in determining cosmological quantities such as the
 Hubble parameter $h(z)$, Om diagnostic $Om(z)$ and the deceleration parameter $q(z)$ \cite{Crossing_C},
 since by reconstructing $d_L(z)$ one can derive $h(z)$, $Om(z)$ and $q(z)$ using: 
\begin{eqnarray}
H(z) =\left[\frac{d}{dz}\left(\frac{d_L(z)}{1+z}\right)\right]^{-1}\\
q(z)=(1+z)\frac{H'(z)}{H(z)}-1\\
Om(z)=\frac{h^2(z)-1}{(1+z)^3-1}
\label{eq:h}
\end{eqnarray}

\noindent where derivatives are respect to the redshift and $h(z)=H(z)/H_0$. Its important
 to note that in applying this method to derive the cosmological quantities in
(\ref{eq:h}) one does not require a prior knowledge 
of the matter density. In fact $Om(z)$ can be used to falsify the $\Lambda$CDM model
without any prior knowledge of $\Omega_{0m}$
 since $Om(z)$ is a constant at all redshifts in $\Lambda$CDM ~\cite{sahni08,zunckel08}. 

\subsection{Comparing Data Sets}

We shall apply our method to compare the consistency of the luminosity distance (obtained
from type Ia supernovae) and the angular size distance (obtained
from X-ray and SZ clusters).
To compare these two data sets one need not reconstruct $h(z)$, $Om(z)$, etc.,
 instead one can simply compare the confidence limits of the Crossing hyperparameters, $C_1$ and $C_2$,
 derived from these different data sets. In order to perform this comparison, we first apply the smoothing method on supernovae data 
to derive $\mu_{Smooth}(z)$ and then use this $\mu_{Smooth}(z)$ 
{\em to fit both luminosity distance data from supernovae as well as
 angular diameter distance data from X-ray and SZ clusters.}
This is done using Chebyshev Crossing functions in (\ref{eq:Cheb})
 so that one derives the Crossing hyperparameters in each case. 
In other words, after deriving $\mu_{Smooth}(z)$ by smoothing supernovae data,
 we fit supernovae data using (\ref{eq:main}) and place constraints on $C_1^{SN}$ and $C_2^{SN}$. 
Since the mean function is derived from supernovae data itself we expect the confidence limits of 
$C_1^{SN}$ and $C_2^{SN}$ to be centred around the $(0,0)$ point in the hyperparameter space.

Next we again use (\ref{eq:main}) (with the same $\mu_{Smooth}(z)$ from supernovae data)  but replace supernovae data by cluster data to determine $C_1^{Cluster}$ and $C_2^{Cluster}$. If the confidence contour of
$C_1^{Cluster}$ and $C_2^{Cluster}$ has a significant overlap with the confidence contour of 
$C_1^{SN}$ and $C_2^{SN}$, then we can conclude that the two data sets (SNIa \& Cluster) are in concordance 
with each other. Otherwise we are observing an inconsistency most probably due to the 
presence of systematics in either of the data sets. The reader should note that in fitting both supernovae and cluster data, and to determine crossing hyperparameters, we use the same smooth mean function, $\mu_{Smooth}(z)$ (Above it was derived from supernovae data).
 In fact the crossing hyperparameters in each case represent the deviation from the mean function 
suggested by the data. If two data sets imply different deviations from a given mean function, $\mu_{Smooth}(z)$, then this might suggest an inconsistency between the different data sets. 

We apply this method to a simulated future SNIa data set assumed to have about 2300 supernovae 
with $0.015<z<1.7$ and $\sigma_{int}=0.13$, and consistent with expectations from the WFIRST survey~\cite{WFIRST}. 
For our second data set we focus on the reasonably good angular diameter distance data expected from future 
SZ and X-ray surveys. Current and upcoming surveys in both radio (SZ), like SPT, ACT or Planck, and X-ray, 
like eROSITA would have large patches of the observed sky in common and hence would be expected to
 have a common ensemble of clusters. If a subset of these clusters have measured temperatures, which will naturally come about during follow-up mass calibrations, then this cluster ensemble can be used to estimate the angular diameter distance to these clusters. The $d_A(z)$ constructed from this dataset will therefore provide a complementary
 probe of cosmology to $d_L(z)$ from SNIa. We construct mock catalogs of such $d_A(z)$ measurements to see the implications for cosmology and systematics. For this purpose, we use clusters that will be potentially found in two all sky surveys, Planck (which has already started discovering clusters using their SZ signatures), and eROSITA (which is the upcoming X-ray survey to be launched in 2014). For Planck, we assume that clusters will be detected over a limiting flux of 300 mJy (at 353 GHz), which will return close to $\sim 2000$ clusters in $\sim 32000$ deg$^2$ of the sky. The higher flux limit means that Planck will detect only the high mass nearby clusters. eROSITA, which is a deeper survey, is modelled to observe clusters over a flux limit of $4\times10^{-14}$ erg cm$^{-2}$s$^{-1}$ in the $\left[ 0.5-2.0 \, \rm{keV} \right]$ band. It will detect $\sim 10^5$ clusters over $\sim 32000$ deg$^2$. All the clusters detected by Planck will be seen by eROSITA.


The clusters that will be jointly found in these two surveys, and hence the constructed $d_A(z)$,
 will depend on the underlying flux-limited cluster redshift distribution $dN/dz$ of these surveys , as well as the redshift distribution of the subset for which the temperature can be measured. We begin by, first, constructing the cluster $dN/dz$ using the corresponding mass-limit, $M_{lim}(z)$, to the above flux-limits. For details see \cite{Khedekar2010b,MM04,Khedekar2012}. In our analysis, we select clusters that lie above the highest of the $M_{lim}(z)$ from either of the surveys.
 This gives the first subset of common clusters. Next, using the fact that we need roughly ten times the photons to measure X-ray temperatures, we calculate the corresponding $M_{lim}(z)$ for ten times the detectable flux. This gives a further subset to the one already realised. Finally, we chose a final subset from these clusters, namely only those that will be large enough to be fit by a simple $\beta$-model; this is possible if the cluster core radius subtends an angular size at the cluster redshift which is at least few times the beam size (typically 2-3 times the eROSITA beam which we take to be $\sim 16$ arc-sec). The corresponding isophotal size is converted to a limiting mass using cluster scaling relations. The redshift-$d_A(z)$ catalog is constructed by integrating the cluster mass function (for our fiducial cosmology) over the highest of these limiting masses. This gives the number of clusters for which we would be able to estimate $d_A(z)$. Finally, the clusters are distributed randomly with a Gaussian scatter of $25\%$ about the $d_A(z)$ from the fiducial cosmology. Since, typically, we expect to measure $d_A(z)$ in the future with better than $20\%$ accuracy, we put a conservative error of $20\%$ on each measure of $d_A(z)$. 
For our mock $d_A(z)$ catalog we get $\sim 1830$ clusters between $0.1<z<1.1$ for a
 fiducial $\Lambda$CDM cosmology. At the end of this paper we also use a fiducial DE model with
 $w=-0.8$ to cross check our proposed method and in that case get more than 2000 clusters up to 
redshifts of $1.2$.

Note that cluster data has the form of $d_A(z)H_0/c$ where $H_0$ is the Hubble constant and $c$ 
is the speed of light, 
whereas Supernovae data involves the distance modulus (\ref{eq:mu}) which is directly related to $d_L(z)$ 
in (\ref{eq:lum_dis}). 
In this context it is useful to 
note that while: (i) $H_0$ acts as a nuisance parameter when dealing with supernovae data 
since it is degenerate with the latter's brightness, 
(ii) $H_0$ is needed to derive $d_A(z)$ from cluster data and this adds an additional element
of  uncertainty to the analysis. 

\begin{itemize}

\item
In the first stage of our analysis we assume that there are no systematics in the data and
that the value of the Hubble parameter is perfectly known. In simulating the data we assume a flat $\Lambda$CDM model with $\Omega_{0m}=0.26$ and $H_0=70 km/sec/Mpc$ and we apply our method of comparison to this
fiducial dark energy model. 
Fig.\ref{fig:nosys} shows the contours of the Crossing hyperparameters derived from SNIa and Cluster
 data and one sees that they have a proper overlap. 
Note that while we have assumed  $\Lambda$CDM to be our fiducial model in simulating the data, similar results would have been obtained using any other fiducial cosmology to generate data. Therefore the results of fig.\ref{fig:nosys} are essentially model independent. We have shown this at a later part of this paper where we did our analysis for another dark energy model with $w=-0.8$ for better clarification.  


\begin{figure*}[!t]
\vspace{-0.8in}
\hspace{0.in}
\includegraphics[scale=0.50, angle=0]{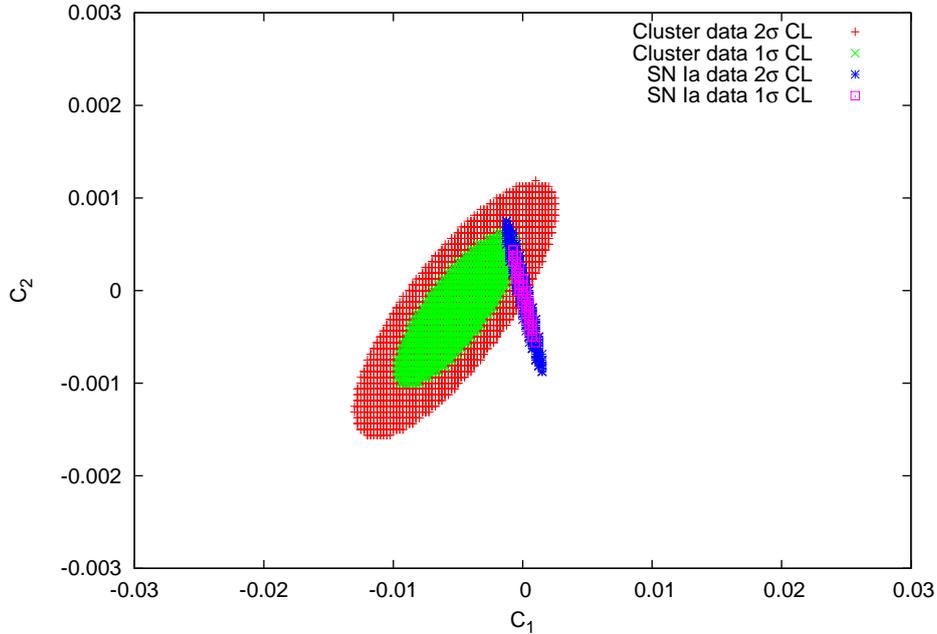}
\vspace{-0.5in}
\caption {\small Confidence contours of the Crossing hyperparameters using simulated future supernovae data from WFIRST survey ($1\sigma$ CL in magenta, $2\sigma$ CL in blue) and simulated future 
 (SZ + X-ray) angular diameter distance data ($1\sigma$ CL in green, $2\sigma$ CL in red). The two data sets are clearly consistent in the absence of
systematics in either of the data. Note that the simulated data are based on a flat $\Lambda$CDM model with $\Omega_{0m}=0.266$ (we have to assume a model to simulate the data in any case) but in our analysis and in the determination of these confidence contours we have not assumed any cosmological model which make our method model independent.
}
\label{fig:nosys}
\end{figure*}

\begin{figure*}[!t]
\centering
\begin{center}
\vspace{1.in}
\centerline{\mbox{\hspace{0.in} \hspace{2.1in}  \hspace{2.1in} }}
$\begin{array}{@{\hspace{-0.3in}}c@{\hspace{0.3in}}c@{\hspace{0.3in}}c}
\multicolumn{1}{l}{\mbox{}} &
\multicolumn{1}{l}{\mbox{}} \\ [-2.8cm]
\hspace{-0.5in}
\includegraphics[scale=0.35, angle=0]{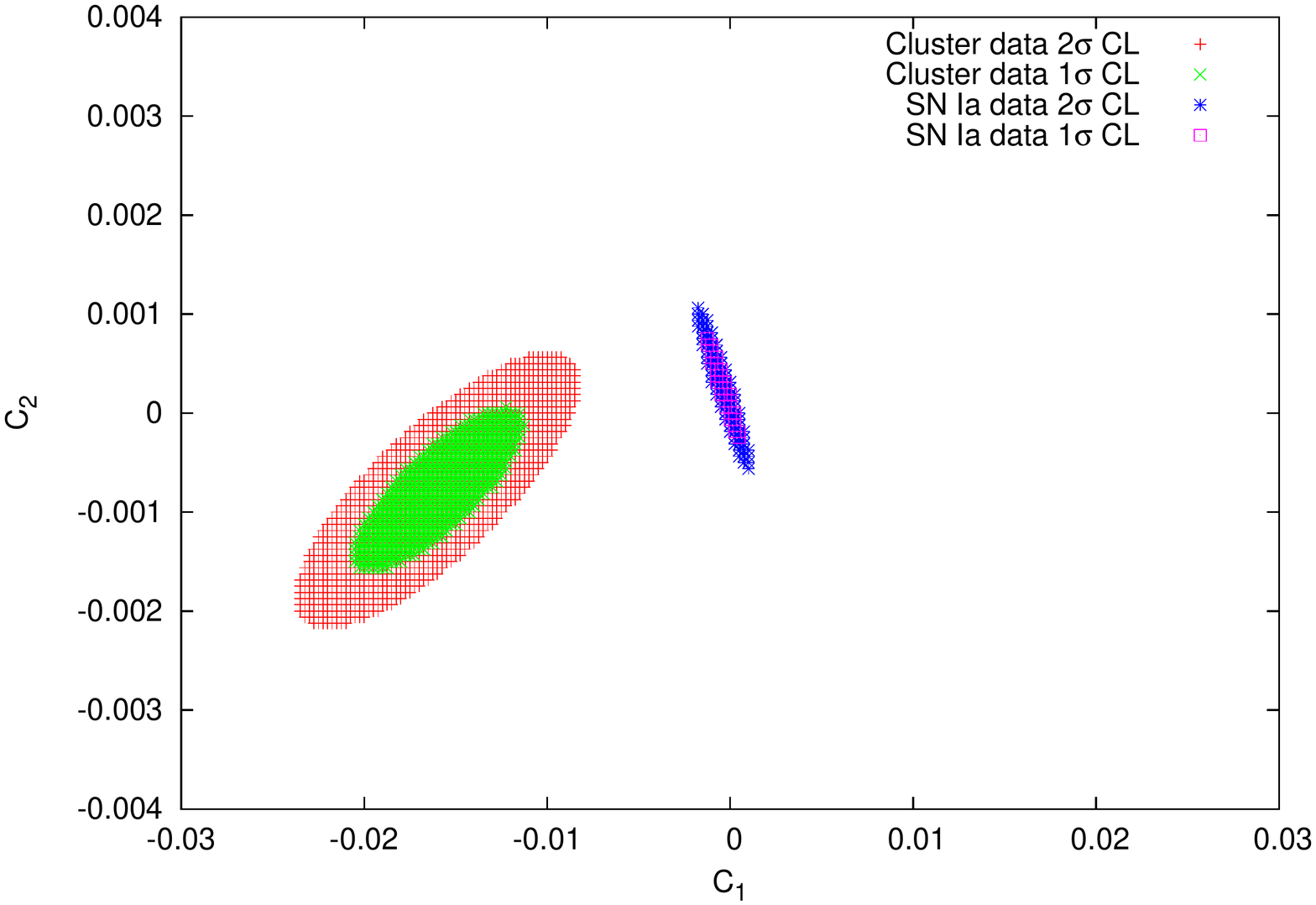}
\hspace{-.5in}
\includegraphics[scale=0.35, angle=0]{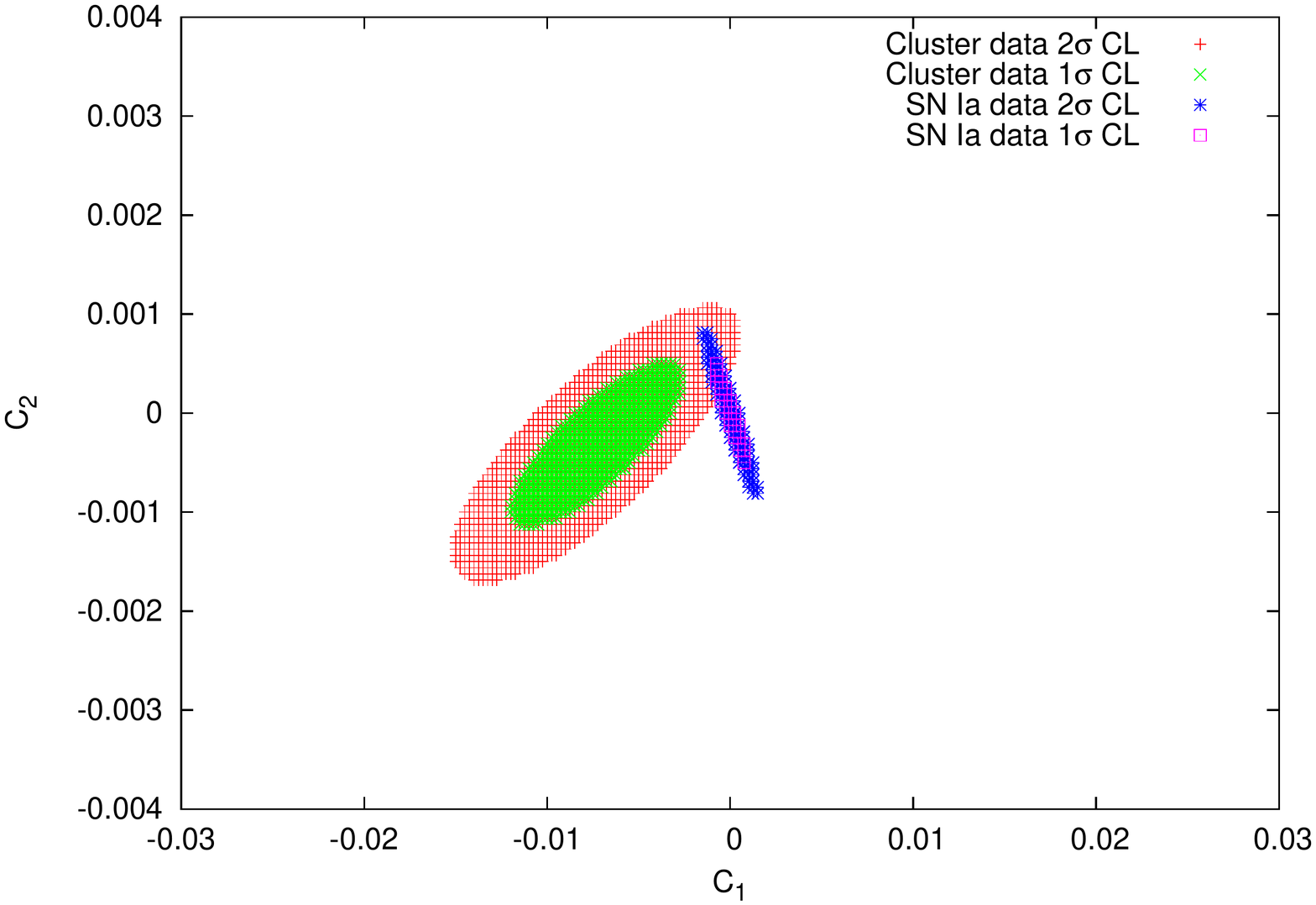}
\hspace{-.1in}
\vspace{0.8in}
\end{array}$
\vspace{-1.3in}
\end{center}
\caption {\small Confidence contours of the Crossing hyperparameters using simulated future supernovae data 
($1\sigma$ CL in magenta, $2\sigma$ CL in blue) and simulated future (SZ + X-ray) angular diameter distance data 
($1\sigma$ CL in green, $2\sigma$ CL in red) after assuming the presence of systematics in the supernovae data. 
The inconsistency between the two data sets is clearly visible both in the left panel (at $> 2\sigma$ CL)
and marginally
so in the right panel (at $1\sigma$ CL) . 
The systematics is reflected in a redshift dependent increase in the distance modulus (1.5) and is
described by (2.7) with $\alpha=0.01$ (left panel) and $\alpha=0.002$ (right panel).}
\label{fig:sys}
\end{figure*}

\begin{figure*}[!t]
\centering
\begin{center}
\vspace{.in}
\centerline{\mbox{\hspace{0.in} \hspace{2.1in}  \hspace{2.1in} }}
$\begin{array}{@{\hspace{-0.3in}}c@{\hspace{0.3in}}c@{\hspace{0.3in}}c}
\multicolumn{1}{l}{\mbox{}} &
\multicolumn{1}{l}{\mbox{}} \\ [-2.8cm]
\hspace{-0.5in}
\includegraphics[scale=0.45, angle=0]{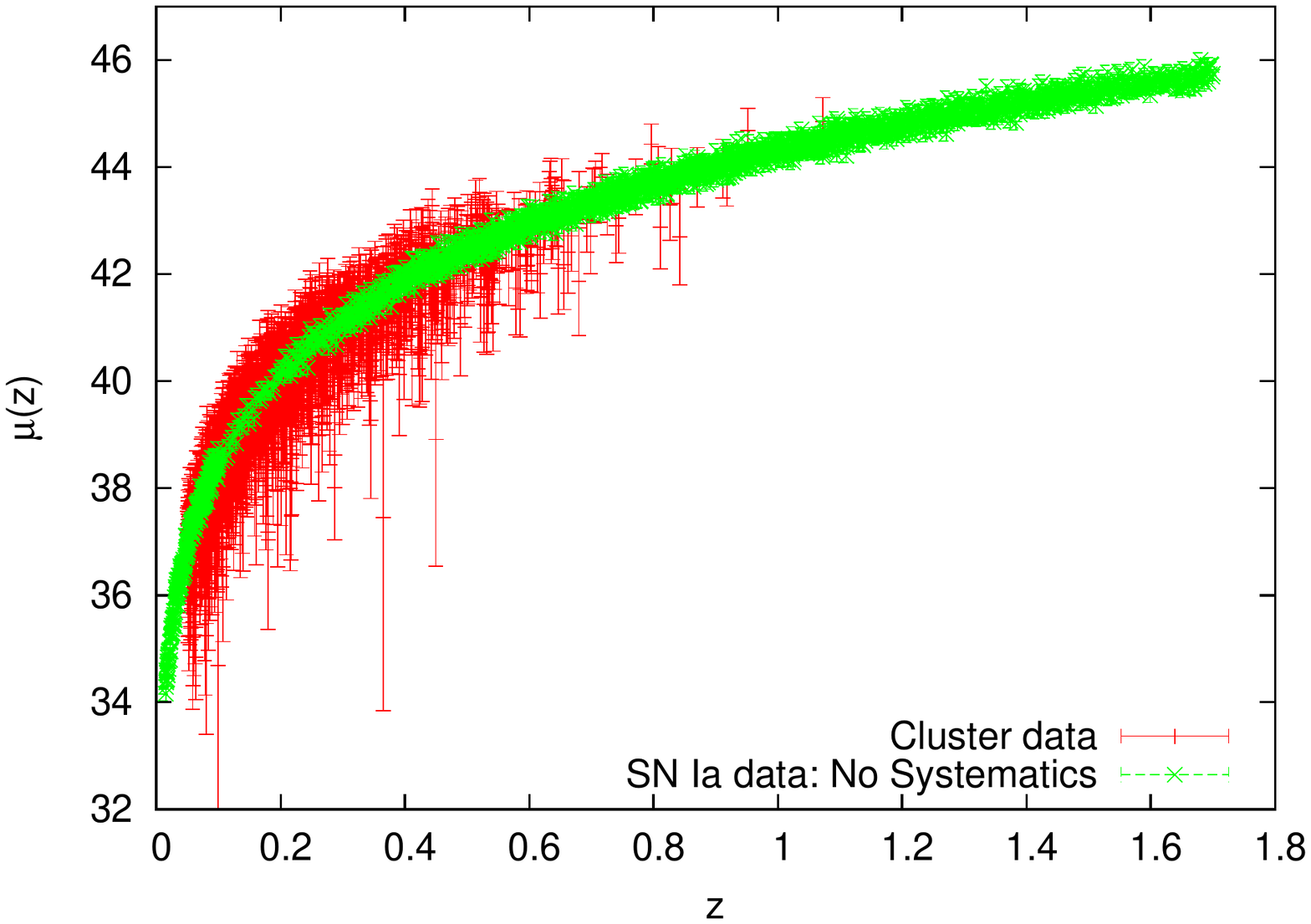}
\hspace{-1.5in}
\includegraphics[scale=0.45, angle=0]{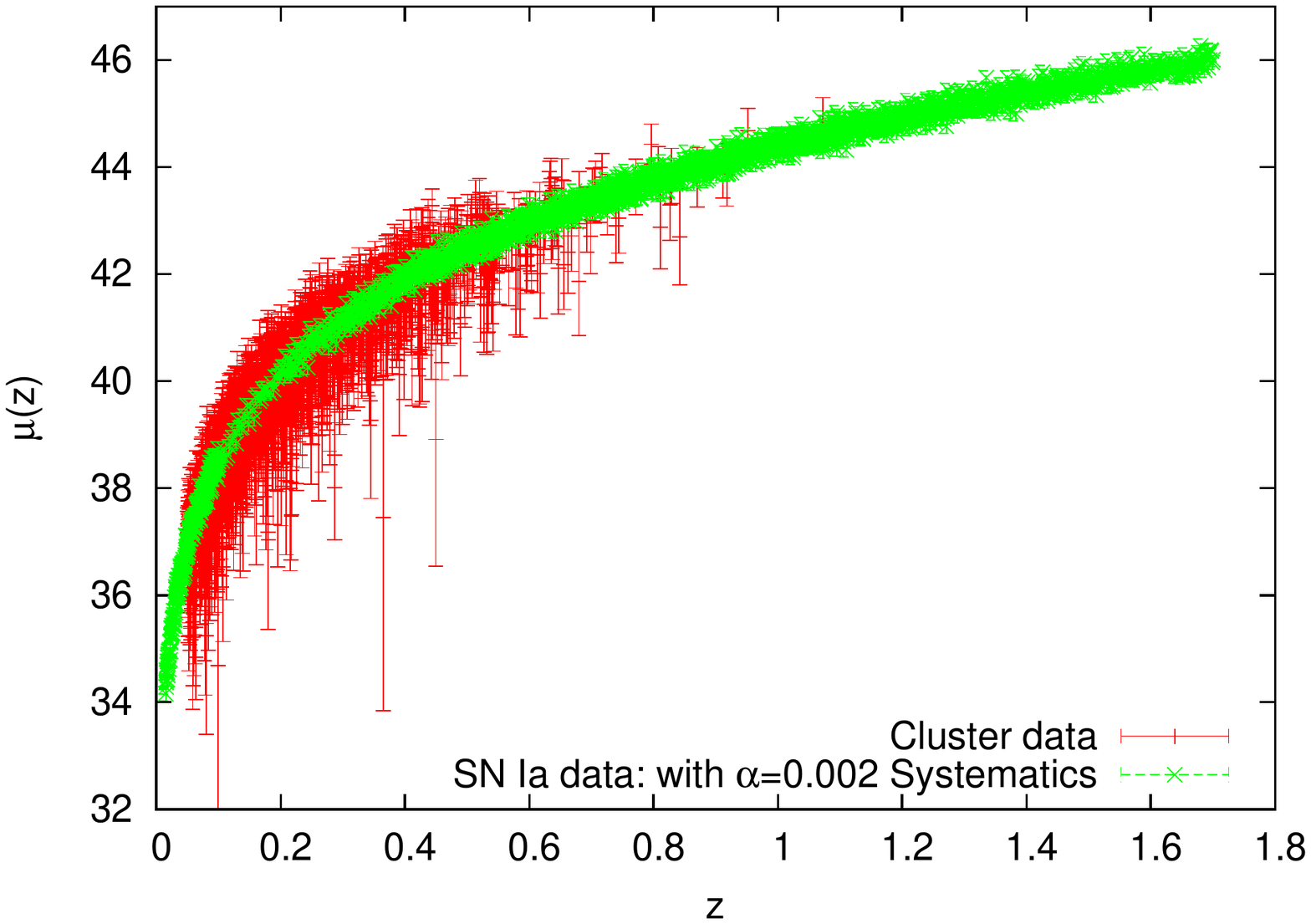}
\hspace{-.1in}
\vspace{0.8in}
\end{array}$
\vspace{-1.3in}
\end{center}
\caption {\small The distance modulus for the SNIa and cluster data is compared in these figures. 
The distance modulus for cluster data has been obtained after applying the cosmic duality relation
(1.4).
The left panel alludes to data with no systematics. The systematic errors in the right panel are
described by $\alpha=0.002$ in (2.7).
It clearly seems difficult to distinguish between the two panels by eye, whereas the crossing statistic
shown in
the right panel of the previous figure accomplishes this.}
\label{fig:data}
\end{figure*}

\begin{figure*}[!t]
\hspace{0.in}
\includegraphics[scale=0.50, angle=0]{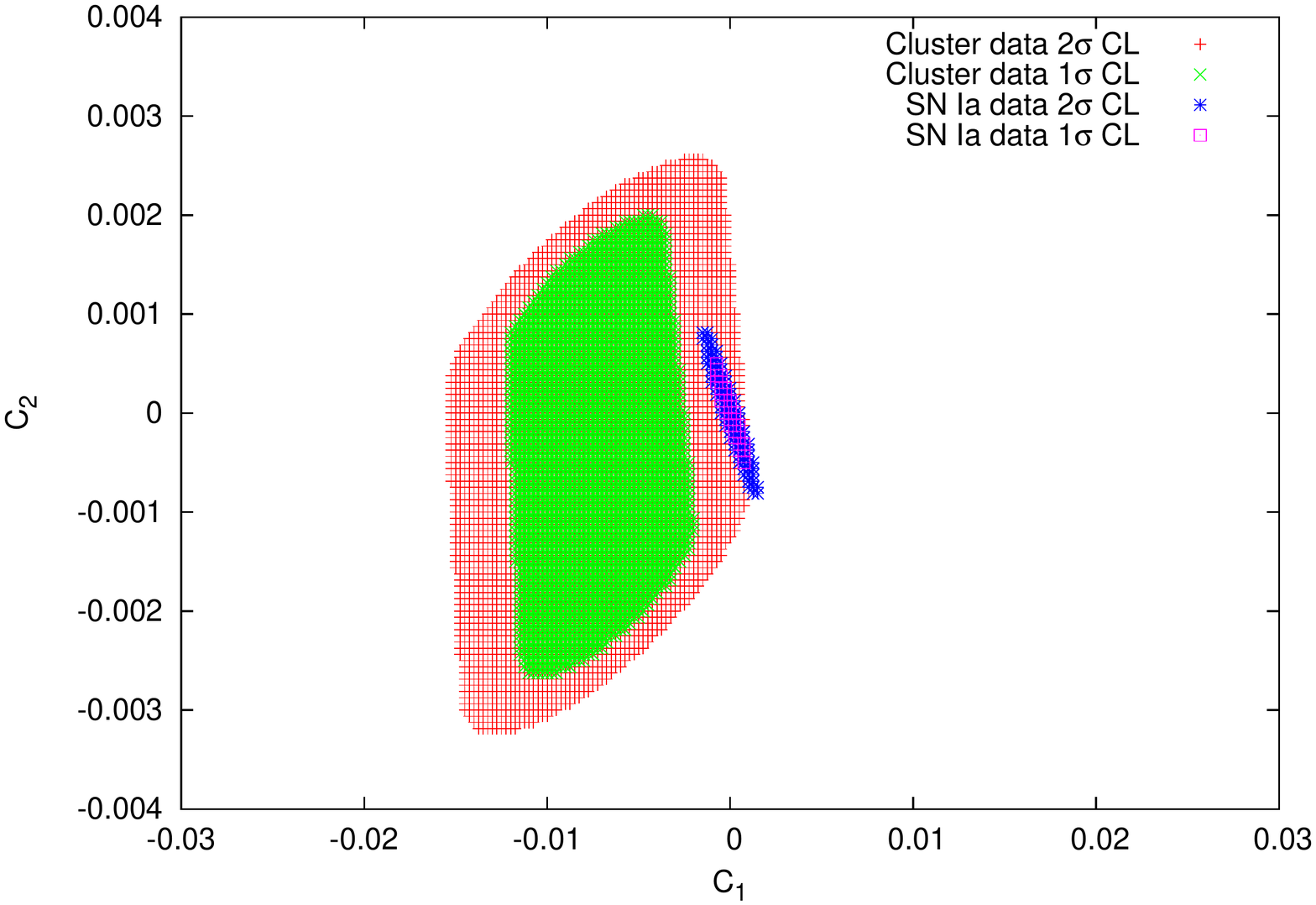}
\vspace{-0.5in}
\caption {\small Confidence contours of the Crossing hyperparameters using simulated future supernovae data 
($1\sigma$ CL in magenta, $2\sigma$ CL in blue) and simulated future SZ angular diameter distance data 
($1\sigma$ CL in green, $2\sigma$ CL in red) assuming 
 the presence of small systematic errors in supernovae data, namely $\alpha=0.002$ in (2.7).
We assume no precise knowledge of $H_0$ choosing instead a flat prior on 
$0.68<h_0<0.72$ ($h_0 = H_0/100~{\rm km/sec/Mpc}$). Interestingly the 
uncertainty in $H_0$ inflates the cluster contour in a direction 
which is orthogonal to that of the SNIa contour
and therefore does not affect our consistency check. 
Consequently one can still pick out the presence of systematics at roughly the $1\sigma$ level even
if the value of $H_0$ is not precisely known and the systematic uncertainties in the data are
exceedingly small. Note that we have not assumed any particular cosmological model for this comparison and analysis.
}
\label{fig:sysH0}
\end{figure*}

\item
Next we assume that the supernovae data set contains systematics which result in the following
 redshift dependent
increase in the value of the distance modulus:

\begin{equation}
\mu_{obs}=\mu_{actual} \times (1+ \alpha z^2)
\label{eq:sys}
\end{equation}

where $\alpha$ quantifies the amount of systematics. The left panel of fig.\ref{fig:sys} shows a
 comparison between the two data sets after assuming $\alpha=0.01$ for the systematics. One clearly sees
 that the two contours are far apart which is indicative of a strong inconsistency between the 
SNIa and Cluster data sets. In the right panel of  fig.\ref{fig:sys} the systematics have been
reduced five-fold to $\alpha=0.002$. Looking at this figure we notice that the two data sets are 
still somewhat
inconsistent since the $1\sigma$ confidence contours do not intersect.
Note that this is hardly the case if one visualises the data themselves, which is what we have done in
fig.\ref{fig:data} in which the data sets corresponding to $\alpha=0$ (left panel)
and $\alpha=0.002$ (right panel) are shown superimposed.
 {\footnote{We have transformed the angular diameter distance data to the distance modulus by 
error propagation after assuming that we know the correct value of the Hubble parameter.}} 
It is clearly difficult to see any visual difference between the left and right panels of this figure,
whereas such a difference is discernable when we compare fig.\ref{fig:sys} (right) with fig.\ref{fig:nosys}.
We believe this
 reflects the strength of our method in discerning the presence of small amounts of systematics in different
data sets. 

\end{itemize}

Note that the analysis thus far assumes that the value of $H_0$ is known perfectly. 
In reality there will be uncertainties in $H_0$ from observations carried out in the local universe. 
In fig.\ref{fig:sysH0} we show the results of a comparison between SNIa and Cluster data after 
assuming $\alpha=0.002$ and allowing the Hubble parameter to have a flat prior in the range $68<H_0<72$. 
This figure shows that an uncertainty in $H_0$ does not seem to affect the detected inconsistency between the two data sets as it inflates the contour volumes in a direction which is perpendicular to
that of overlap (between the contours). This is good news since we are still able to test for data systematics
 even without  a precise knowledge of $H_0$, thereby mimicking the current observational situation.  

\begin{figure*}[!t]
\hspace{0.in}
\includegraphics[scale=0.50, angle=0]{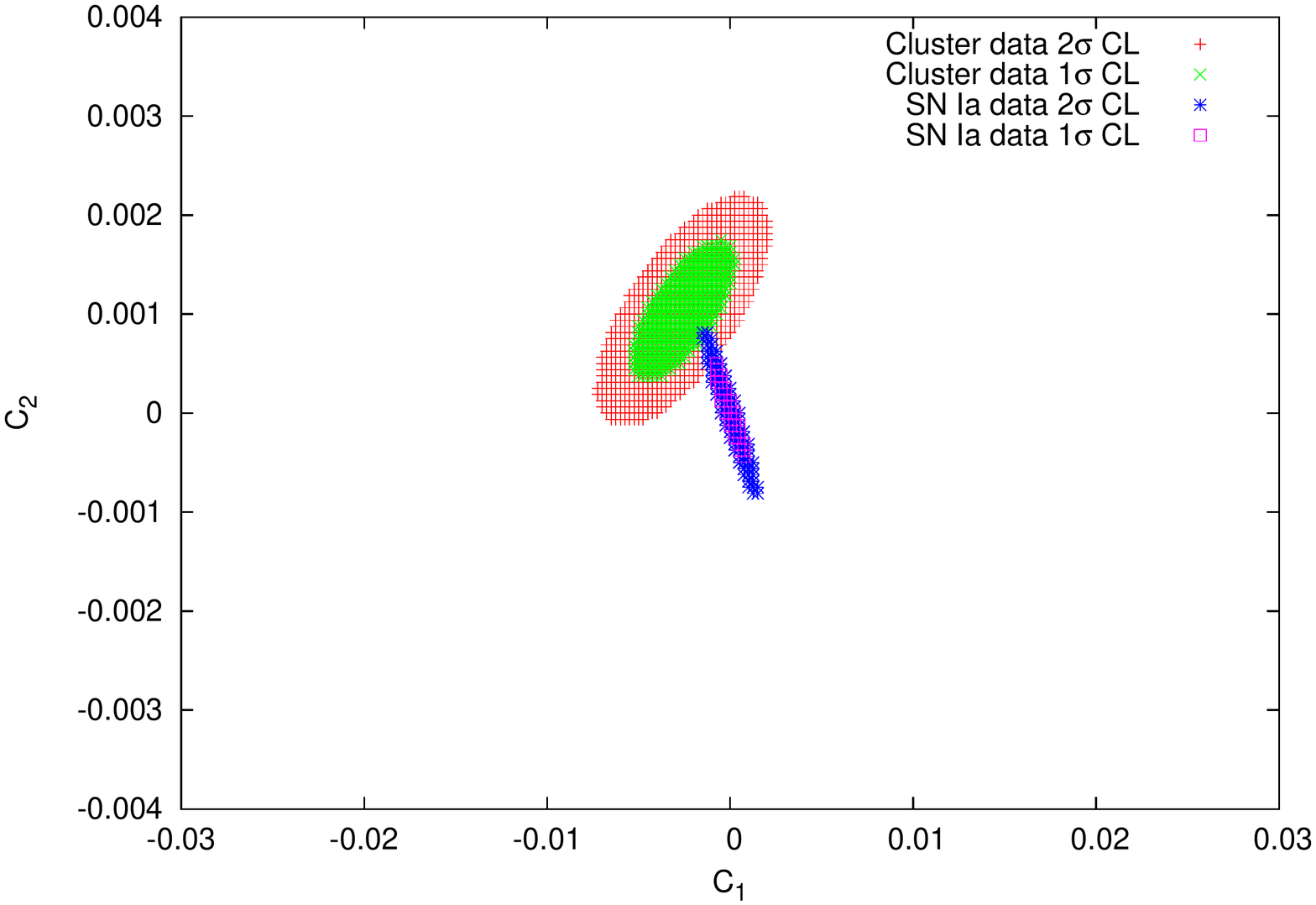}
\vspace{-0.5in}
\caption {\small Confidence contours of the Crossing hyperparameters using simulated future supernovae data from WFIRST survey ($1\sigma$ CL in magenta, $2\sigma$ CL in blue) and simulated future 
(SZ + X-ray) angular diameter distance data ($1\sigma$ CL in green, $2\sigma$ CL in red) for a quiescence dark energy model with $w=-0.8$. 
The SNIa data set has an inbuilt systematic encoded in a $\alpha=0.002$ magnitude shift in (2.7). 
This leads to an inconsistency between the confidence contours corresponding to the two data sets as seen
in the figure. Although the above figure shows results for DE with an unevolving equation of state,
 a different dark energy model could have served our purpose equally well  
since our approach is not sensitive to the nature of dark energy.
} 
\label{fig:wmp8}
\end{figure*}

As a final check, we test our method using a different fiducial model in order to make sure that 
it can be applied to different models of dark energy. We simulate the data for dark energy
with $w=-0.8$ and our results are presented in fig.\ref{fig:wmp8}. This figure assumes 
systematic magnitude shift errors for supernovae similar to the previous case, namely $\alpha=0.002$. 
Again the confidence contours corresponding to the SNIa and Cluster data sets are well distinguished,
 demonstrating that our approach in identifying the presence of systematics 
is independent of the background cosmological model.

\section{Conclusion}                        
\label{concl}

In this paper we present a robust and easy to use method of comparing different cosmological data sets 
in order to search for possible systematics in them. 
It has been shown earlier that combining the smoothing method and Crossings Statistic results in an approach which can be easily used to reconstruct the expansion history of the universe and the properties of dark energy without setting any priors on cosmological parameters \cite{Crossing_C}. In this paper we have used the idea of 
Bayesian interpretation of Crossing statistic \cite{Crossing_B} to test the consistency of two important cosmological data sets, 
namely the luminosity distance to type Ia supernovae,
 and the angular diameter distance to galaxy clusters, in order to search
 for possible systematics in either of the data sets. We have shown that our method can identify
 the presence of systematics (such as a small magnitude shift in supernovae data) 
in a model independent manner and even in
the absence of a precise knowledge of the Hubble constant, $H_0$. The method developed by us succeeds in discerning the presence of systematics at the sub-percent level in data sets expected to become
available in the near future. The method we used in this paper can be generally used to compare different data sets in a model independent manner and this work is in fact one application of the method to compare angular diameter distance data with luminosity distance data assuming the cosmic duality relation.

\acknowledgments{A.S. thanks Eric Linder for useful discussions. A.S. acknowledge the Max Planck Society (MPG), the Korea Ministry of Education, Science and Technology (MEST), Gyeongsangbuk-Do and Pohang City for the support of the Independent Junior Research Groups at the Asia Pacific Center for Theoretical Physics (APCTP). S.M. would like to thank Satej Khedekar for collaborations on cluster data. A.A.S. acknowledges RESCEU hospitality as a visiting professor. He was also partially supported by the grant RFBR 11-02-00643 and by the Scientific Programme $\Pi$-21 "Astronomy" of the Russian Academy of Sciences.}

\appendix

\end{document}